\documentclass[journal=jpccck,manuscript=article,layout=twocolumn]{achemso}

\usepackage{chemformula} % Formula subscripts using \ch{}
\usepackage[T1]{fontenc} % Use modern font encodings

\usepackage[utf8]{inputenc}
\usepackage[english]{babel}
\usepackage{graphicx}% Include figure files
\usepackage{dcolumn}% Align table columns on decimal point
\usepackage{bm}% bold math
\usepackage{color}
\usepackage{enumerate}
\usepackage{hyperref}% add hypertext capabilities
\title{Exploring librational pathways with on-the-fly machine-learning force fields: Methylammonium molecules in MAPb$X_3$ ($X=$ I, Br, Cl) perovskites}
\author{Menno Bokdam}
\email{m.bokdam@utwente.nl}
\affiliation{University of Twente, Faculty of Science and Technology and MESA+ Institute for Nanotechnology, P.O. Box 217, 7500 AE Enschede, The Netherlands}
\alsoaffiliation{University of Vienna, Faculty of Physics and Center for Computational Materials Sciences, A-1090 Vienna, Austria}
\author{Jonathan Lahnsteiner}
\affiliation{University of Twente, Faculty of Science and Technology and MESA+ Institute for Nanotechnology, P.O. Box 217, 7500 AE Enschede, The Netherlands}
\author{D.D. Sarma}
\affiliation{Solid State and Structural Chemistry Unit, Indian Institute of Science, Bengaluru 560012, India}

\date{\today}% It is always \toda, today,
\let\oldmaketitle\maketitle
\let\maketitle\relax

\begin{document}

\twocolumn[
\begin{@twocolumnfalse}
\oldmaketitle
\begin{abstract}
Two seemingly similar crystal structures of the low-temperature ($\sim$100~K) MAPb$X_3$ ($X$=I,Br,Cl) perovskites, but with different relative Methylammonium (MA) ordering, have appeared as representatives of this orthorhombic phase. Distinguishing them by X-ray diffraction experiments is difficult and conventional first-principles based molecular-dynamics approaches are often too computationally intensive to be feasible. Therefore, to determine the thermodynamically stable structure, we use a recently introduced on-the-fly Machine-Learning Force Field method, which reduces the computation time from years to days. The molecules exhibit a large degree of anharmonic motion depending on temperature: i.e. rattling, twisting and tumbling. We observe the crystal's ’\textit{librational pathways}’ while slowly heating it in isothermal-isobaric simulations. Marked differences in the thermal evolution of structural parameters allow us to determine the real structure of the system via a comparison with experimentally determined crystal structures. 
\end{abstract}
\end{@twocolumnfalse}
]

\begin{tocentry}
\textbf{TOC Graphic}\newline{}

\includegraphics[width=8.25cm]{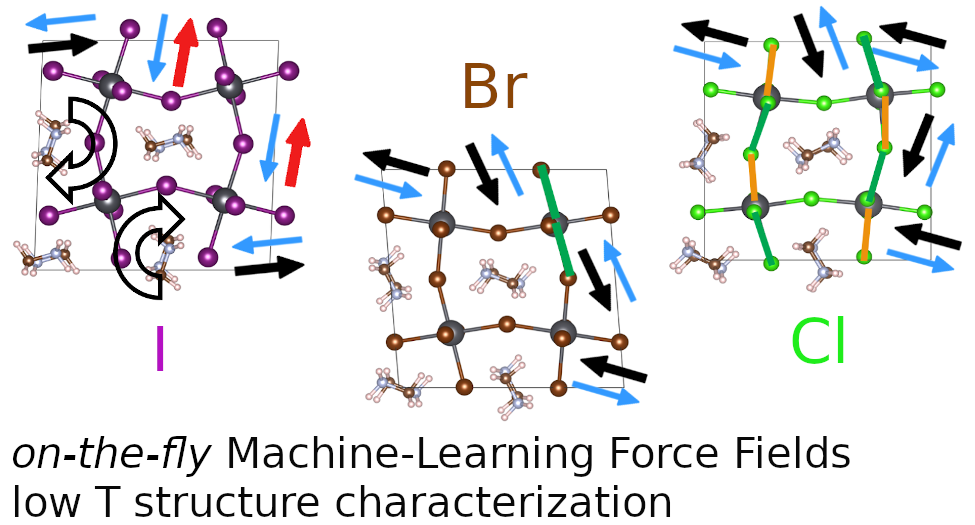}
\end{tocentry}

\section{Introduction}

The crystal structure of hybrid halide-perovskites is a topic of study that has surfaced several times in the last four decades. X-ray powder diffraction experiments of Weber~\textit{et.al.} on Methylammonium(MA)-Pb$X_3$ with halogens $X$ = \{I, Br, Cl\}, have established a high temperature cubic phase for all $X$\cite{Weber:zfn78}. A perovskite structure is formed by Pb$X_6$ corner-sharing octahedra enclosing the MA molecules. In later years, the low temperature phases and librational modes of the MA molecule at various temperatures were studied\cite{Wasylishen:ssc85,Poglitsch:jcp87,Onoda-Yamamuro:jpcs90,Kawamura:jpsj02,Mashiyama:jkps07}. The development of high efficiency solar cells based on these perovskites sparked a rival of interest in its structure characterization\cite{Stoumpos:ic13,Baikie:jmca:13,Weller:chc15,Govinda:jpcl16}. It was shown by high-quality powder neutron-diffraction experiments\cite{Weller:chc15} that the low-temperature orthorhombic phase of MAPbI$_3$ actually belongs to the $Pnma$ space-group. This is the space group that was also determined for MAPbBr$_3$\cite{Swainson:jssc03} and MAPbCl$_3$\cite{Chi:jssc05}. The potential for opto-electronic applications raised new questions that are all (in)directly related to the atomic structure: the effect of MA rotation on charge dynamics\cite{Gelvez-Rueda:jpcc16,Fabini:jacs17}, dynamic or permanent deformations of the Pb$X_6$ octahedra\cite{Beecher:acsel16,Page:acie16,Bernasconi:acsenl17,Bernasconi:jpcc18}, the extent of electron-phonon coupling\cite{Wright:natc16,Sender:math16,Neukirch:nanol16} and the Rashba effect\cite{Stroppa:natc14,Azarhoosh:aplm16,Leppert:jpcl16,Etienne:jpcl16,Hutter:natm17,Frohna:natc18}, to name a few. Considerable progress has been made, but consensus has not always been achieved. This is in part the result of differences in interpretation of the local microscopic structure. The disorder, be it static or dynamic, of the molecular C-N axes is a well-known problem for diffraction techniques, that makes it difficult to determine their precise orientation\cite{Bernasconi:jpcc18,Wiedemann:jpcl21}. First-principles (FP) methods such as density functional theory (DFT) have shown to be very useful for the determination of crystal structure by augmenting the experimentally resolved inorganic framework with the ordering of the molecules\cite{Filippetti:jpcc14,Quarti:com14,Mattoni:jpcc15,Brivio:prb15,Bakulin:jpcl15,Lahnsteiner:prb16,Carignano:jpcc2017,Lahnsteiner:prm18,Maheshwari:jpcc19}. However, even though commonly used density functional approximations have the required chemical accuracy\cite{Bokdam:prl17}, their computational complexity prohibits the large length \&{} time scale molecular dynamics (MD) calculations necessary to resolve the free energy landscape and thereby the finite temperature crystal structure\cite{Mattoni:jpcc15}. We will use the \textit{on-the-fly} Machine-Learning Force Field (MLFF) method\cite{Jinnouchi:prl19,Jinnouchi:prb19}, which makes it possible to explore the full diversity of atomic structures while going through the entropy-driven phase transformations in hybrid perovskites. This method substantially reduces the computational cost while retaining near-FP accuracy. Recently, it has been shown to be capable to resolve the orthorhombic-tetragonal (Ort-Tet) and tetragonal-cubic (Tet-Cub) phase transitions in MAPbI$_3$ and the inorganic halide perovskites CsPb$X_3$ in good agreement with experiment\cite{Jinnouchi:prl19}. Furthermore, it can be systematically extended to describe mixed MA$_x$FA$_{1-x}$PbI$_3$ perovskites under isothermal-isobaric conditions\cite{Grueninger:jpcc21}.

\begin{figure}[!t]
    \begin{center}
    \includegraphics[width=\columnwidth ,clip=true]{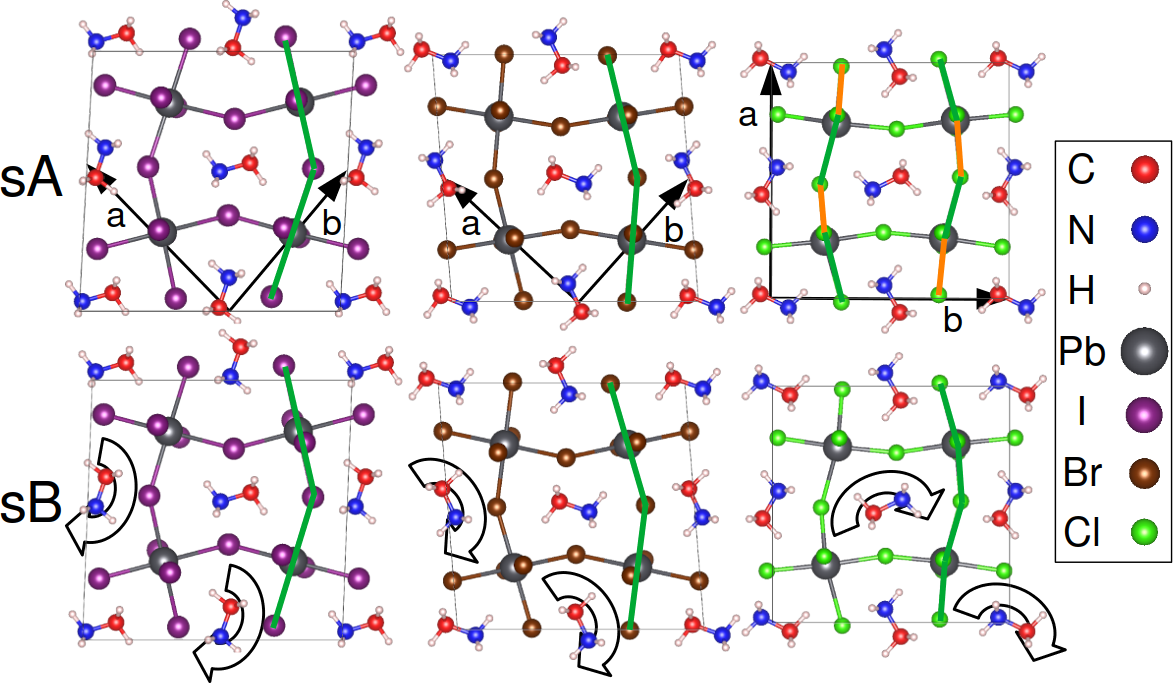}
    \end{center}
   \caption{Initial low-temperature crystal structures of MAPb$X_3$, with molecules in the sA and sB arrangement. Lattice vectors and Pb,$X$ atom coordinates are adapted from Refs.~\cite{Baikie:jmca:13},\cite{Swainson:jssc03} and \cite{Chi:jssc05}. Only one of the two layers of molecules in the $ab$-plane is shown. Molecules in the second layer (into the paper) are anti-parallel to the first. sB is created by rotating half of the molecules of sA by 180$^{\rm o}$ in the ab-plane. The distortions of the PbCl$_6$ octahedra are indicated in the top right figure by the long (green) and short (orange) Pb-Cl bonds. }
\label{fig:1}
\end{figure}

%-----------------------------------------------------------------------

The starting point of our search for the low-temperature ($\sim$100~K) orthorhombic structure of MAPb$X_3$ are two seemingly similar, but distinctly different structures: sA and sB. They have the same lattice vectors and inorganic coordinates, but a different molecular ordering pattern as sketched in Figure~\ref{fig:1}. We have labeled the lattice vectors such that the molecules lie in the $ab$-plane. sB is created out of sA by an in-plane rotation of half of the molecules by 180$^{\rm o}$ as indicated by the curved arrows. Note that in both arrangements the neighboring molecules in the $c$-direction (not shown in the figure) are anti-parallel. These structures have been prepared in a $2\times2\times2$ supercell, such that it accommodates the ($\sqrt{2}a_{\rm p},\sqrt{2}a_{\rm p},2a_{\rm p}$) basis of $X$ = I, Br  as well as the ($2a_{\rm p},2a_{\rm p},2a_{\rm p}$) basis of $X$ = Cl, where $a_{\rm p}$ is the pseudo-cubic lattice constant of the parent $P{m\overline{3}m}$ cell \cite{Chi:jssc05}. The molecules in the often referenced experimental ($Pnma$) structures for $X$ = I, Br and Cl of Refs~\cite{Baikie:jmca:13,Swainson:jssc03} and \cite{Chi:jssc05} are arranged as in the sB, sA and sA configuration, respectively. Other experimental works did not distinguish between these two arrangements\cite{Stoumpos:ic13,Whitfield:sr16}, since refinement of the model structure by permuting N and C with respect to the measured diffraction spectra does not lead to significant improvements of the fit\cite{Bernasconi:jpcc18}. To date even for the extensively studied MAPbI$_3$ perovskite, different studies report opposite arrangements: sA\cite{Brivio:prb15,Omer:epjb18,Leveillee:prb19} and sB\cite{Filippetti:jpcc14,Filip:prb14,Lee:com16,Lahnsteiner:prb16}. This is unexpected, because if we focus only on the dipole moment of the MA molecule and compute the total electrostatic energy in the point-dipole approximation, then the sB pattern is clearly favored.  This pattern shows a closer resemblance to the `head-tails' groundstate of a point-dipole model\cite{Lahnsteiner:prb19}.\newline{}

In this work, we will analyze the `\textit{librational pathways}' of the MA molecules and Pb$X_6$ octahedra in MAPb$X_3$, and use them to identify the most representative low-temperature  orthorhombic structure. We sample structures of the crystal by slowly heating up two plausible low-temperature structures (sA and sB) in isothermal-isobaric MD simulations. Structures on the explored pathways through the structural phase space are thermodynamically linked to the starting configuration and result in marked differences in lattice and order parameters that are compared to temperature dependent diffraction studies.

\section{Computational Details}

The DFT calculations are performed with the projector augmented wave method\cite{Blochl:prb94b} as implemented in the VASP code\cite{Kresse:prb96,Kresse:cms96} using the meta-gradient corrected SCAN\cite{Sun:prl15} density functional approximation (DFA), which has shown good performance when compared to high quality many-body perturbation theory reference calculations\cite{Bokdam:prl17}. A plane-wave basis with a cutoff of 350~eV, Gaussian smearing with a width of 10~meV and 4 (I,Br) or 8 (Cl) k-points of the $\Gamma$-centered 2$\times$2$\times$2 Monkhorst-Pack grid are set, which suffice to obtain the required accuracy of the calculations\cite{Lahnsteiner:prm18}. The computed lattice parameters as function of temperature should (qualitatively) agree with experiment over the whole temperature range. Therefore, by not limiting the study to a 0~K DFT based relaxation of the internal energy, biases related to the chosen DFA can be detected\cite{Bokdam:prl17}. Before starting the MD simulations the starting structures of Fig.~\ref{fig:1} were shortly relaxed by a conjugate gradient algorithm. \newline

MLFFs are trained during MD simulations with VASP, based on calculated total energies, forces and stress tensors for automatically (on-the-fly) selected structures in the isothermal–isobaric ensemble. This approach is described in
detail in Refs. ~\cite{Jinnouchi:prl19,Jinnouchi:prb19}. In short, a Bayesian error estimation of the predicted forces is used to select either DFT or MLFF forces to propagate the structure in time ($t_n\rightarrow t_{n+1}$). Whenever the predicted errors exceed the threshold, a new reference structure is picked up, a DFT calculation is performed and the coefficients of the MLFF are re-optimized. In Figs.~\ref{fig:NDFT}(c,f) a 'density-of-states' like function of the temperature (note, equivalent to simulation time) shows when in the training MD most DFT calculations were performed. It is calculated by: $\rho_{\rm FP}(T)=\sum_{i=0}^{\rm N_{ref}}\delta_i(T-T_i)$, where $\delta(T)$ is a Lorentzian function. This function is normalized to the total number of DFT reference structures picked up in training, ${\rm N_{ref}}=\int \rho_{\rm FP}(T)dT$. The automatically picked up reference structures form a minimal training database (containing total energies, forces, stress tensors and atomic coordinates) that is well spread over the available structural phase space. We have shared this database via the \href{https://doi.org/10.4121/14710323.v1}{4TU.DataBase} repository\cite{Bokdam:4TUDATA21} to encourage development of ML potentials based on minimalistic datasets.

A variant of the GAP-SOAP\cite{Bartok:prl10,Bartok:prb13} method is
used as a descriptor of the local atomic configuration around each atom. Within a cutoff of 7~\AA{} a two-body radial probability distribution $\rho^{(2)}_i(r)$ is build, as well as three-body angular distribution $\rho^{(3)}_i(r,s,\theta)$ within a cutoff of 4~\AA{}. The atomic coordinates are smeared in the distributions by placing Gaussians with a width of 0.5~\AA{}. The obtained distributions are projected 
on a finite basis set of spherical Bessel functions multiplied with spherical harmonics. The bessel functions are of the order 6 and 7 for the radial and angular part, respectively. Only the angular part has a  maximal angular momentum of $l_{max}=6$. The
coefficients of the projections are gathered in the descriptor vector $\mathbf{X}_i$. A kernel-based
regression method\cite{Bishop:book06} is applied to map the descriptor to a local atomic energy. The similarity between two local configurations is calculated by a polynomial kernel function: $
   K(\mathbf{X}_i,\mathbf{X}_{i_B})=\nicefrac{1}{2}(\mathbf{X}^{(2)}_i
	     \cdot{}\mathbf{X}^{(2)}_{i_B})+\nicefrac{1}{2}(\mathbf{X}^{(3)}_i\cdot{}
	        \mathbf{X}^{(3)}_{i_B})^4.$
\newline

%-------------------------------------------------------------------------
\begin{figure}
    \begin{center}
    \includegraphics[width=\columnwidth ,clip=true]{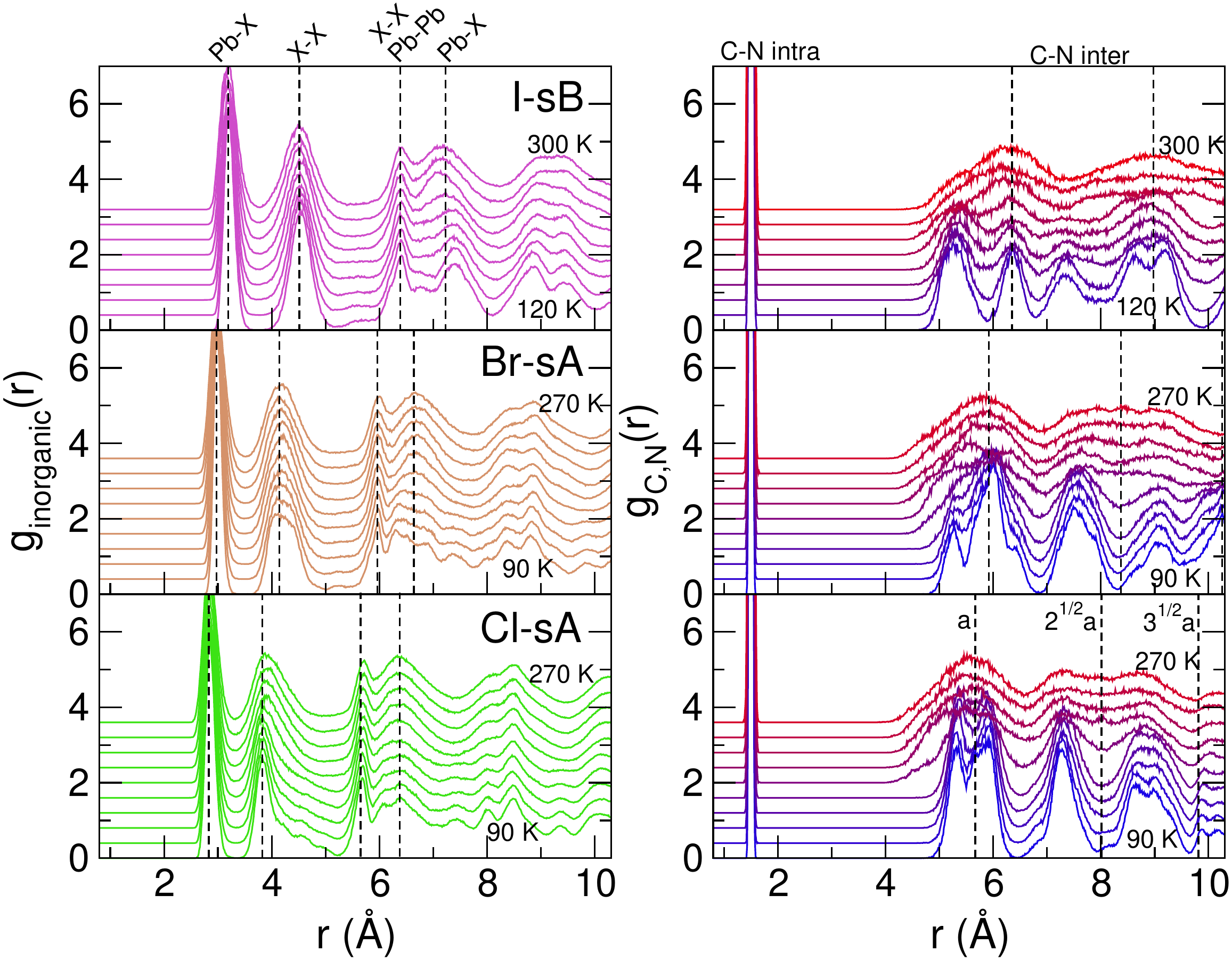}
    \end{center}
   \caption{Combined pair distribution functions of the inorganic (Pb,$X$) components compared to $g_{\rm C,N}(r)$ in the MAPb$X_3$ ($X$= I, Br and Cl) perovskites for increasing temperatures with steps of 20~K. The crystals librate from the initial structures I-sB, Br-sA and Cl-sA. Vertical dashed lines indicate typical bond lengths (left) and neighbor distances based on the lattice constant $a$ in the cubic phase (right).}
\label{fig:rdfintro}
\end{figure}
%-------------------------------------------------------------------------

%--------------------------------------------------------

\begin{figure}[!t]
    \begin{center}
    \includegraphics[width=.9\columnwidth ,clip=true]{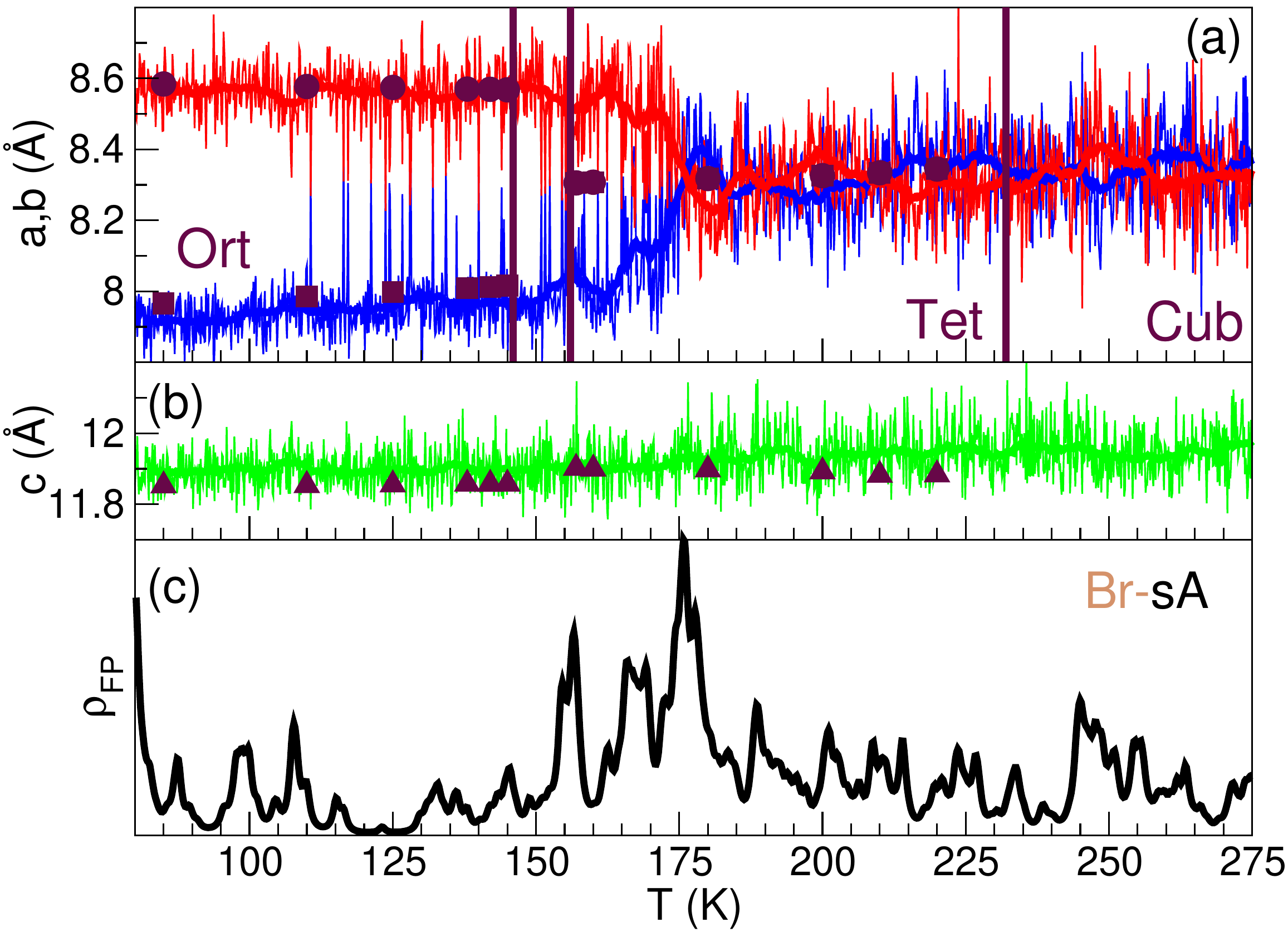}
    \includegraphics[width=.9\columnwidth ,clip=true]{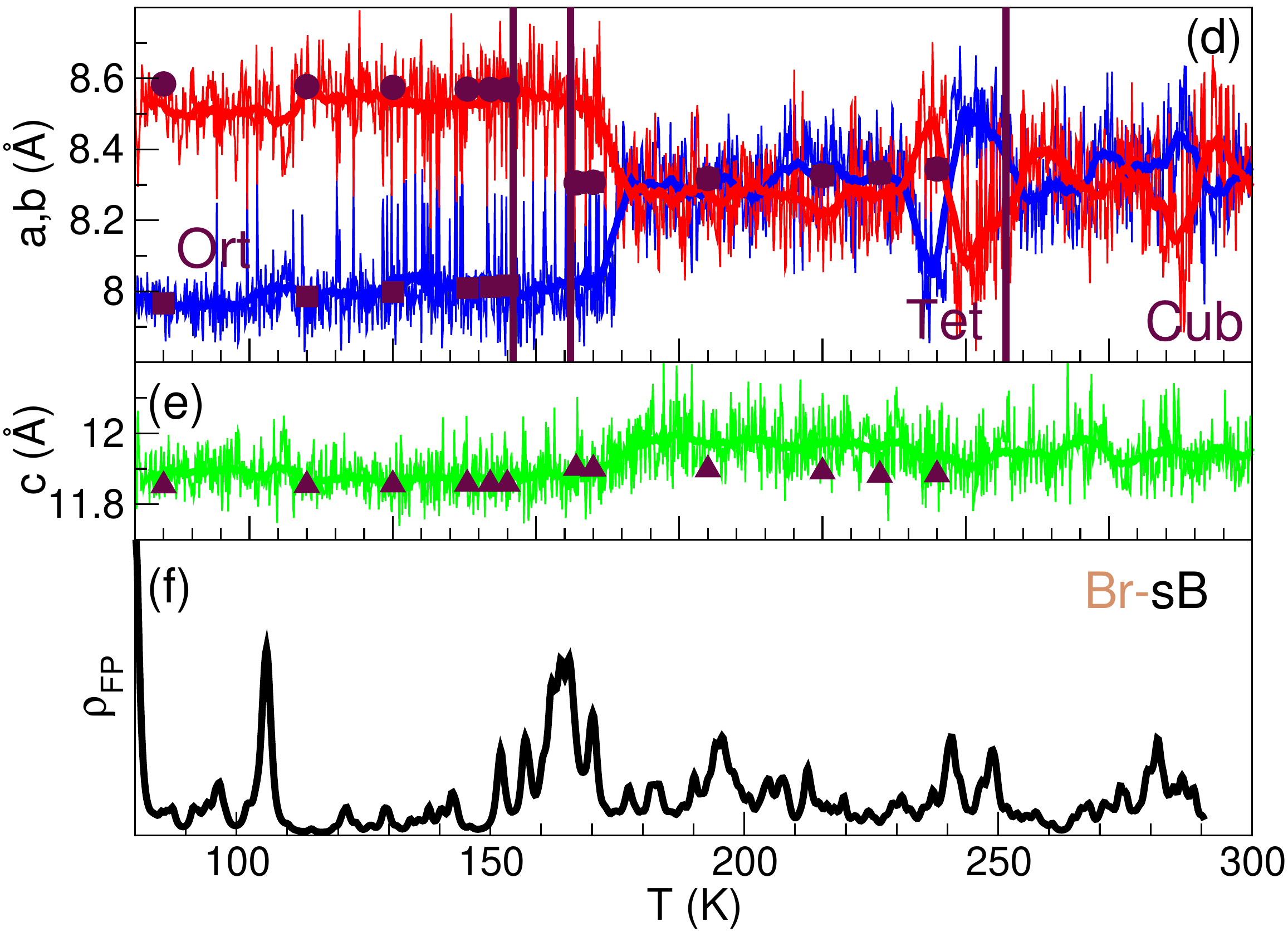}
    \end{center}
   \caption{On-the-fly heating MD at $\frac{2}{3}\frac{\rm K}{\rm ps}$ of MAPbBr$_3$ starting from the sA and sB arrangement. The $a,b$ (a,d) and $c$ (b,e) lattice parameters (blue/red/green) of the orthorhombic system and their running averages (thick lines, window size 5~K). Experimental reference data from Ref.~\cite{Swainson:jssc03}: lattice parameters (symbols) and transition temperatures (thick vertical lines). (c,f) The density $\rho_{\rm FP}(T)$ of performed DFT calculations as function of temperature (black line).}
\label{fig:NDFT}
\end{figure}
%-------------------------------------------------------------------------
\section{Results \&{} Discussion}

To introduce the librational pathways of MAPb$X_3$ we will illustrate them by weighted sums of pair distribution functions (PDFs) in Figure~\ref{fig:rdfintro}. The PDF for the atom types $\alpha$ and $\beta$ is defined as
\begin{equation*}
 g_{\alpha,\beta}(r)=\frac{1}{4\pi r^2dr}\frac{{\rm V}}{{\rm N}_\alpha {\rm N}_\beta}\left <\sum_{i\in\alpha}\sum_{\substack{j\in\beta \\(i\ne j)}}\delta(|\mathbf{r}-\mathbf{r}_{ij}|)\right > ,
\end{equation*}
where $\mathbf{r}_i$ and $\mathbf{r}_j$ are the coordinates of the ${\rm N}_\alpha$ and ${\rm N}_\beta$ atoms, $\mathbf{r}_{ij}=\mathbf{r}_i-\mathbf{r}_j$, $\rm V$ is the volume of the simulation box, and $\left< .\right >$ denotes the ensemble average. In $g_{\rm inorganic}(r)$, the pairs of framework components Pb-$X$,$X$-$X$,Pb-Pb are included, and only the C-N pairs of the MA molecules are included in $g_{\rm C,N}(r)$. For all halides $X$, we see that $g_{\rm inorganic}(r)$ retains most of its structure throughout the whole temperature range, and that $g_{\rm C,N}(r)$ shows a transition whereby part of the order is lost. The intra-molecular part ($r\approx{}1.5$~\AA) remains intact, but the inter-molecular pairs show dual peaks merging into a single broad peak centered around the nearest-neighbor distances of the consecutive cubic unit cells, ie. $\sqrt{1}a_p,\sqrt{2}a_p,\sqrt{3}a_p$. This is the result of the unfreezing of the molecules whereby they reorient ($C_4$) rapidly\cite{Onoda-Yamamuro:jpcs90}. Around room temperature, neighboring MA molecules are still dynamically correlated to their neighbors\cite{Lahnsteiner:prb16}. The differences in the DSF between the halides $X$ are, among other things, related to different Pb-$X$ bond lengths and to the relative orientation of the molecules in the low temperature phase. As we will show hereafter, the thermodynamically stable molecular configurations at low temperature (sA or sB) depends on the halide type $X$.

The starting structures (Fig.~\ref{fig:1}) are slowly heated at a rate of $\frac{2}{3}\frac{\rm K}{\rm ps}$ using DFT based MD with a Langevin thermostat and a time step of 2~fs. The PDFs at different temperatures have been obtained by partitioning the resulting trajectory in parts of equal length. In the $NpT$ ensemble all lattice degrees of freedom are allowed to change as shown for MAPbBr$_3$ in Figure~\ref{fig:NDFT}. The two plots correspond to the heating trajectories starting from the sA and sB structures. Thermal fluctuations in structural parameters are smoothed by applying running averages. To accelerate the MD a MLFF is trained on-the-fly as described in Refs.~\cite{Jinnouchi:prl19,Jinnouchi:prb19}. The algorithm switches between MLFF and DFT forces based on the predicted error of the MLFF. Structural reference configurations to train the MLFF are automatically picked up and by construction lie outside the already 'learned' part of the phase space. This can be seen by the sharp increase in the density of first-principles calculations ($\rho_{\rm FP}(T)$) shown in Figs.~\ref{fig:NDFT}(c,f). The on-the-fly algorithm decides to do a large number of DFT calculations in the region between 150 and 170~K, ie. when the system undergoes the Ort-Tet phase transition. The total number of DFT reference structures picked up in training,  ${\rm N_{ref}}=\int_{80}^{280} \rho_{\rm FP}(T)dT$, is ${\rm N_{ref}^{sA}}=933$ and ${\rm N_{ref}^{sB}}=1074$. This transition temperature is expected to be retarded, because the system is out of thermal equilibrium as a result of the still considerable heating rate. Even though, the agreement with the exp. lattice parameters shown by the symbols in Fig.~\ref{fig:NDFT} is remarkable. We show that the structural transformation and the related librational pathways of the molecules and octahedra (see Supplementary Movies) are accurately described in the on-the-fly heating MD. This opens up the possibility to explore many different perovskites, because only $\sim$1.000 out of the total of 150.000 MD steps per heating run were DFT calculations. This reduces the compute time from years to days.\newline{}

%-------------------------------------------------------------------------
\begin{figure}
    \begin{center}
    \includegraphics[width=\columnwidth ,clip=true]{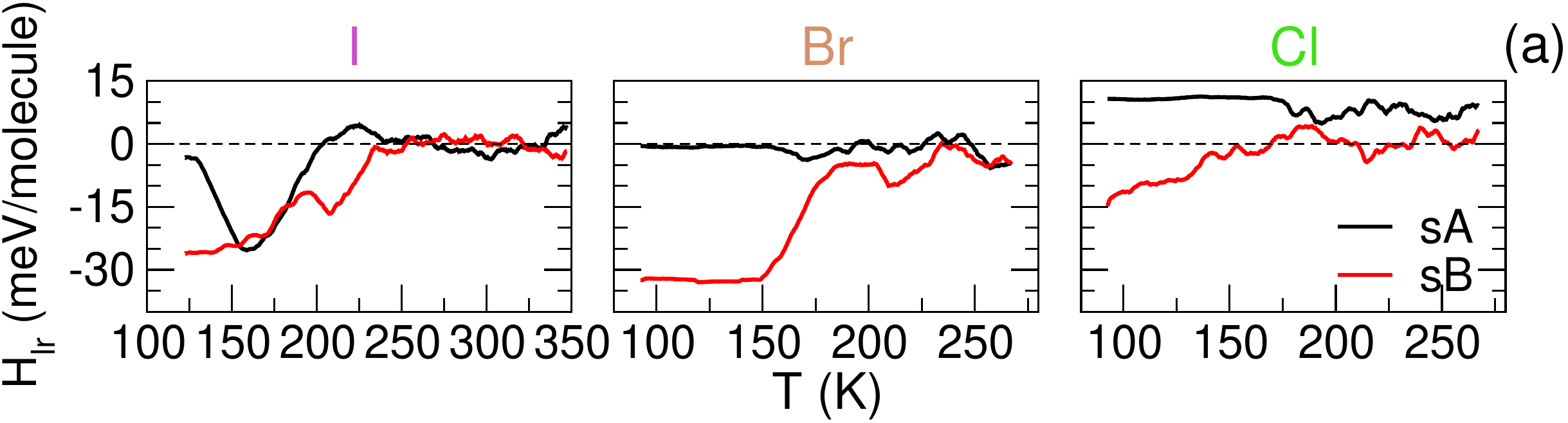}
    \includegraphics[width=\columnwidth ,clip=true]{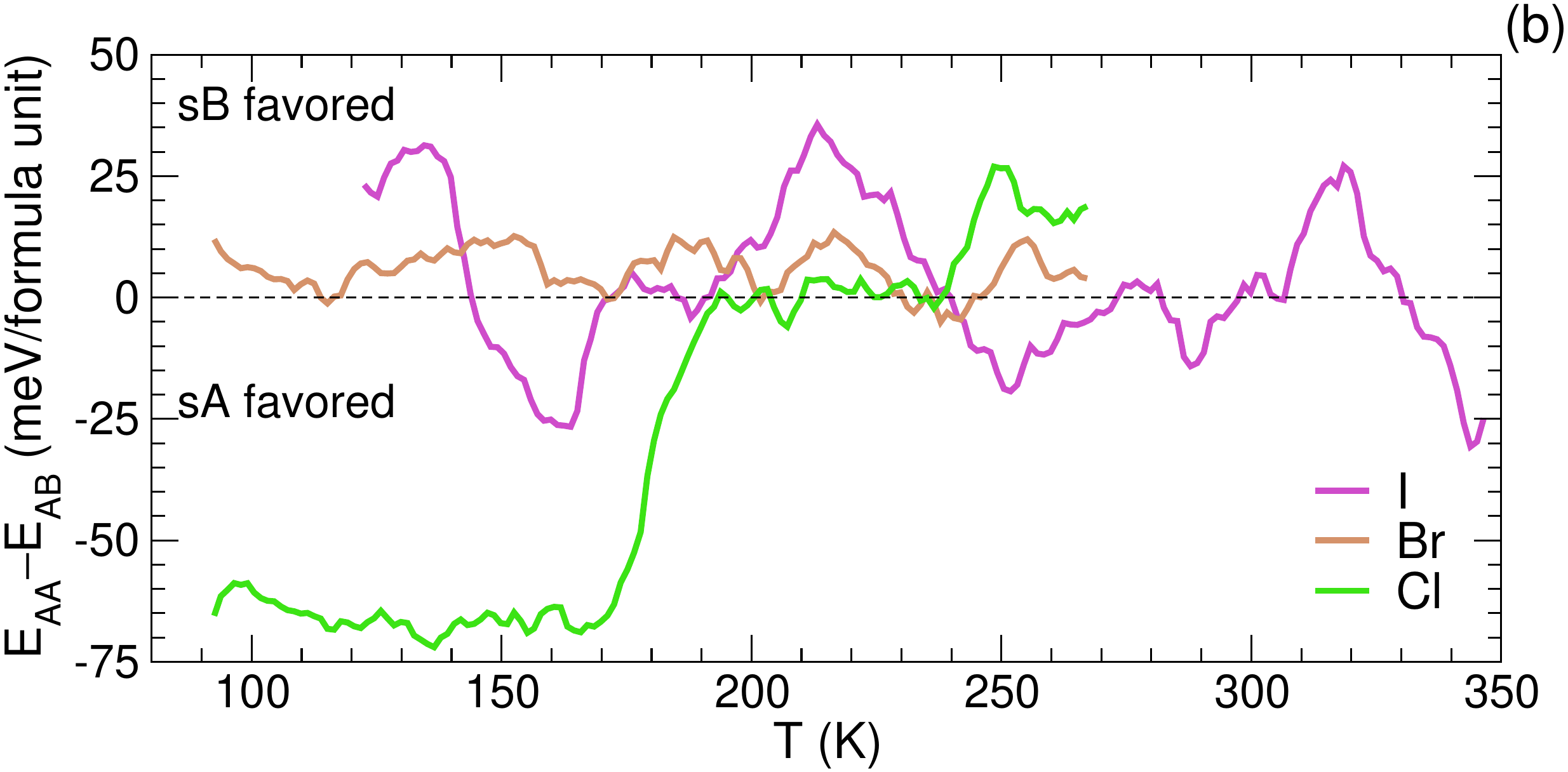}
    \end{center}
   \caption{Energies during the heating trajectories of the MAPb$X_3$ perovskites. (a) Electrostatic energy of the molecular dipoles in the point-dipole approximation ($H_{\rm lr}$) and (b) DFT internal energy differences $(\Delta E)$. (running averages over 25~K)}
\label{fig:2}
\end{figure}
%-------------------------------------------------------------------------
\subsection{Librational pathway analysis}

%-------------------------------------------------------------------------
\begin{figure*}[!t]
    \begin{center}
    \includegraphics[height=6cm ,clip=true]{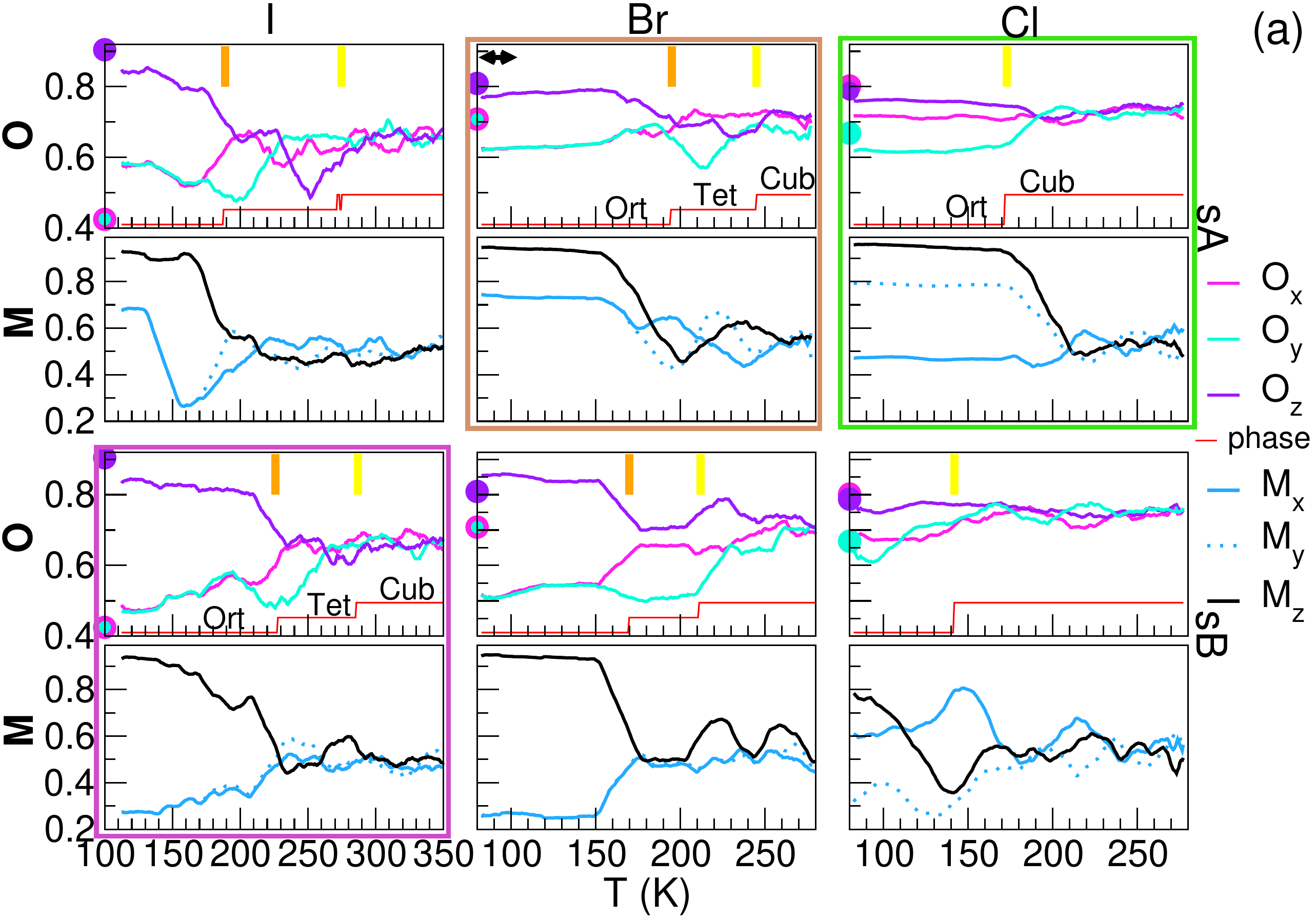}
    \includegraphics[height=6cm,clip=true]{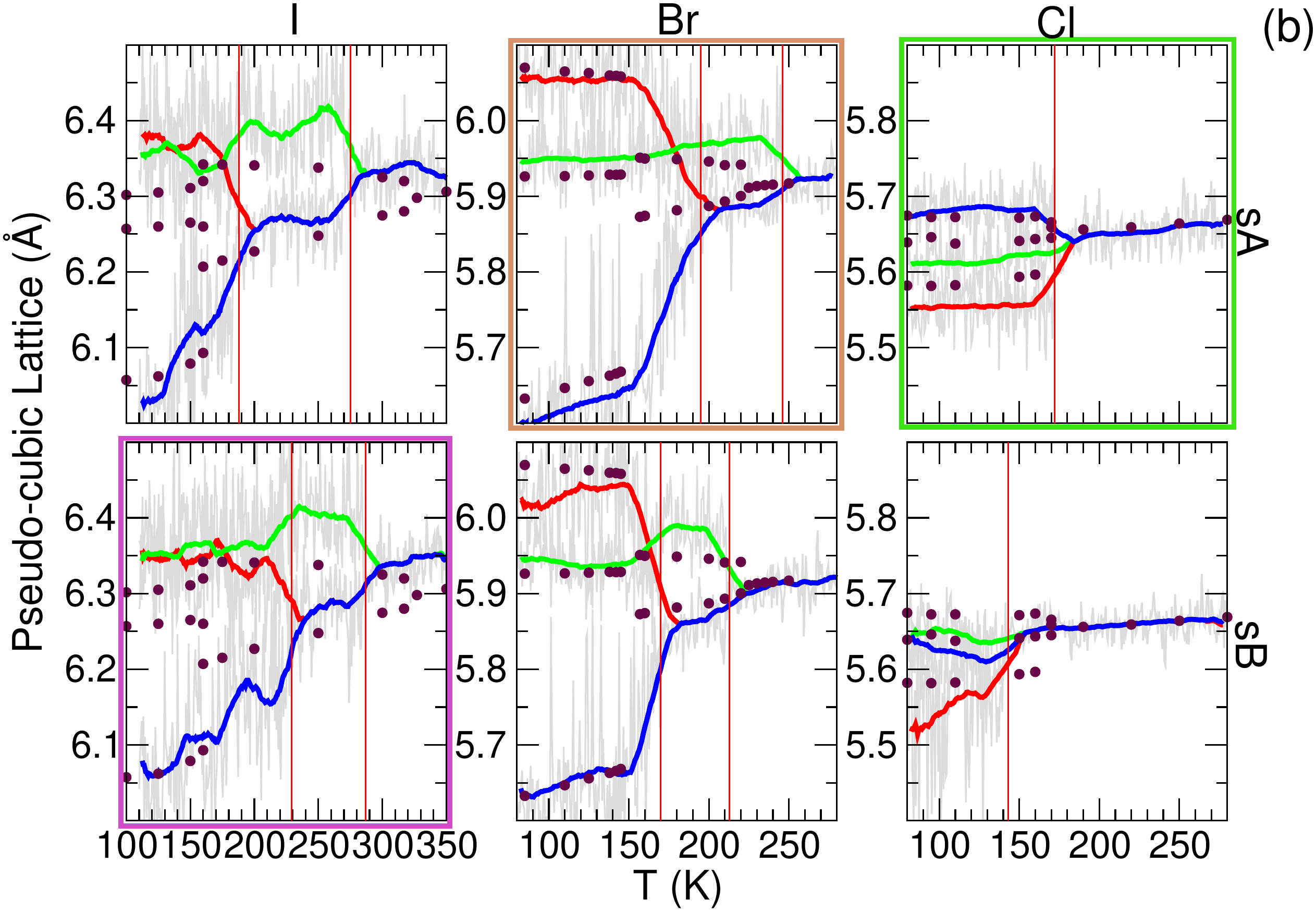}
    \end{center}
   \caption{Heating trajectories of MAPb$X_3$ starting in the sA and sB configuration. The (a) structural order parameters for the octahedra ($\mathbf{O}$) and molecules ($\mathbf{M}$). The red line shows the classified perovskite crystal based on $\mathbf{O}$. (b) The pseudo-cubic lattice constants refined in the classified phase. (running averages over 25~K) The solid circles represent (a) the $\mathbf{O}$ values for the experimental low-temperature structures\cite{Baikie:jmca:13,Swainson:jssc03,Chi:jssc05} and (b) the experimental lattice parameters\cite{Whitfield:sr16,Swainson:jssc03,Chi:jssc05}. Note that the exp. Br $\mathbf{O}$ lies outside the scale of the graph (a) at 11~K.}
\label{fig:3}
\end{figure*}
%-------------------------------------------------------------------------

Figure~\ref{fig:NDFT} indicates that, apart from a small step in the $c$ lattice constant in the sB case, both trajectories qualitatively and quantitatively agree with the experimental data. This means that the lattice parameters alone provides insufficient information to select either the sA or the sB as the thermodynamically stable low-temperature structure. Therefore, we analyze the structural motif presented by the atomic coordinates in the six heating trajectories. First, we extract solely the orientation of the molecular C-N axes in time. The total electrostatic energy ($H_{\rm lr}$) corresponding to the dipole moments of the molecules is calculated\cite{Lahnsteiner:prb19}. In short, this point-dipole model assumes a fixed dipole moment on all molecules, no screening and includes all dipole-dipole interactions up to the third nearest-neighbor. In Figure~\ref{fig:2}(a) $H_{\rm lr}$ is plotted as function of temperature for the three halides starting from the sA (black lines) and sB (red lines) configurations. The sB configuration is clearly lower in energy. With increasing temperature the molecules flip/re-order and the stable arrangement is broken down, eventually leading to a disordered state with $H_{\rm lr}=0$\cite{Govinda:jpcl17}. For $X$ = I the sA pattern flips to the stable sB pattern before reaching the Ort-Tet phase transition, which shows that the initial structure was out of equilibrium. However, for $X$ = Br and Cl the pattern remains largely frozen-in until the phase transition. The sA configuration can only be stable at low-T if either a potential energy contribution arising from the inorganic framework or the entropy compensates this internal energy difference.

The DFT/MLFF calculated internal energy ($E$) is shown in Figure~\ref{fig:2}(b) as the energy difference ($\Delta{}E=E_{\rm sA}-E_{\rm sB}$) between the heating trajectory starting with the sA and the sB structure. At low temperature and ambient pressure, the volume and entropy contributions to the Gibbs free energy are small, and a sizable positive/negative $\Delta{}E$ would indicate that the sB/sA configuration is favored, respectively. For MAPbCl$_3$ it is clear that the sA structure is favored even though its electrostatic energy in the dipole model was unfavored. Above $\sim 175$~K the difference between the two initial configurations has been lifted by thermally induced structural rearrangements. For MAPbI$_3$ and MAPbBr$_3$ the situation is less clear, at low temperature $\Delta{}E$ is positive, however it is of the size of the fluctuations. Even for the fully DFT relaxed (0~K) structures $\Delta{}E$ values are small: 33, $3$ and $-72$ meV/formula-unit for I, Br and Cl, respectively. Increasing the precision of the calculation, by doubling the k-point grid density and applying the Tetrahedon method\cite{Blochl:prb94a} results in 33, $-6$ and $-75$ meV/f.u., respectively. This indicates that, especially for MAPbBr$_3$, we cannot distinguish sA from sB based on the internal energy alone.\newline{}

The changes of the structural motif as a function of the temperature are compared in Figure~\ref{fig:3} and in the Supplementary Movies. Order parameters describing the inter-octahedral ($\mathbf{O}$) and inter-molecular ($\mathbf{M}$) order are shown for all heating trajectories in Fig.~\ref{fig:3}(a). The $\mathbf{M}$/$\mathbf{O}$ order parameters are based on the dot products of $X$-Pb-$X$/C-N connection vectors located on nearest neighboring sites.
A detailed description can be found in Refs.~\cite{Lahnsteiner:prb19,Jinnouchi:prl19}. Looking at $\mathbf{M}$ for I-sA, we see that the molecular ordering pattern rapidly changes starting from  $\mathbf{M}_{\rm 120~K}=\left(0.9,0.7,0.7\right)$ and transforming to the same order observed for I-sB $\mathbf{M}_{\rm 160~K}=\left(0.9,0.3,0.3\right)$, in agreement with the previously seen change in $H_{\rm lr}$. At the same time the inter-octahedral order parameter for the inorganic framework, $\mathbf{O}$, changes only little. The breakdown of the initial molecular order occurs at a lower temperature in the Br case and coincides with the orthorhombic-tetragonal phase transformation. This shows that the sA and sB molecular order are more energetically competitive in orthorhombic $X$ = Br than in I, and that an additional measure is required to determine the stable arrangement. The $\mathbf{O}$ values for the experimental low temperature structures indicated by the circles in Fig.~\ref{fig:3}(a) provide this measure. Based on their comparison to the simulation the Br-sA is the stable low-temperature orthorhombic structure.\newline{}

To automatically classify the instantaneous crystallographic phase of all structures within the MD trajectory, we applied a new approach based on the $\mathbf{O}$ order parameter. This allows, for example, to still assign the orthorhombic phase to a structure that is strained in a box with $a\approx{}b\approx{}c$. Whenever the variance of the components of $\mathbf{O}$ is below a threshold it is classified as Cub, and otherwise it is Ort or Tet. We can then differentiate between the last two by counting the number (1$\rightarrow{}$Ort, 2$\rightarrow{}$Tet) of components larger than their mean value. The red line in Fig.~\ref{fig:3}(a) shows the result of this phase classification. Note that no Tet phase in MAPbCl$_3$ is recognized. This could be caused by a very small temperature window in which the Tet phase is stable, or that the $c/a$ ratio is too small to be noticed in the supercell.

Lattice parameters as function of temperature are shown in Figure~\ref{fig:3}(b). The parameters are refined in the unit cell corresponding to the classified phase and converted to pseudo-cubic lattice parameters ($a_p$). Experimentally obtained parameters are shown by the circles. Surprisingly, the plot for MAPbBr$_3$ and MAPbCl$_3$ show good to very good agreement with experiment. Especially for MAPbI$_3$ we notice the effect of our limited MD setup, suppressing the Tet phase on the $T$-axes. From our previous study we know that this can be improved with a lower heating rate and by applying a larger $4\times4\times4$ supercell. Still, it is noteworthy that, under the same computational settings, the Ort-Tet phase transition becomes more retarded going from Cl, Br to I.

Combining all the above presented analysis leads to the favored initial configurations, I: sB, Br: sA, Cl: sA. These configurations are highlighted by the colored rectangles in Fig.~\ref{fig:3}.

%-------------------------------------------------------------------------
\begin{figure}[!b]
    \begin{center}
    \includegraphics[width=\columnwidth ,clip=true]{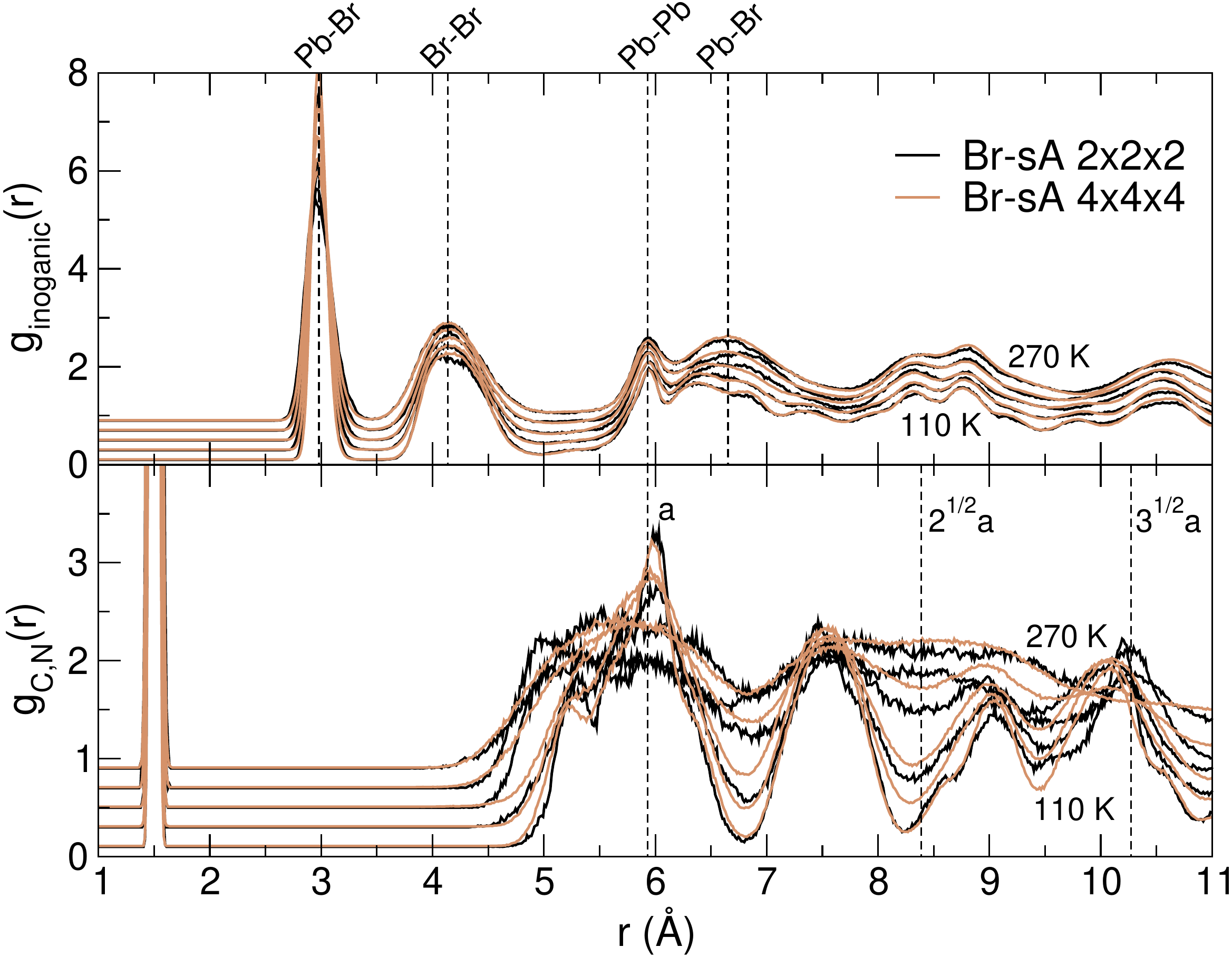}
    \end{center}
   \caption{Influence of supercell dimension. Pair distribution functions of the inorganic (Pb,Br) components (top) and the C,N pairs (bottom) in MAPbBr$_3$ obtained during on-the-fly heating MD in a $2\times2\times2$ and $4\times4\times4$ supercell. Increasing temperatures shown with steps of 40~K.}
\label{fig:sizedep}
\end{figure}
%-------------------------------------------------------------------------

\subsection{Training system size dependence}
The PDFs in Fig.~\ref{fig:rdfintro} are plotted beyond half of the simulation box width, since the $2\times2\times2$ supercell has an average width of $2a_{\rm p}$. They are computed in $4\times4\times4$ supercells, which are created by replicating the original $2\times2\times2$ cell, and enables us to sample the PDF on the $[0,2a_{\rm p}]$ domain. These results are compared to a training run performed with a $4\times4\times4$ supercell. Specifically, we have made a test for MAPbBr$_3$ starting in the sA configuration, and applied the same on-the-fly training directly on the $4\times4\times4$ supercell. In this supercell, the k-points of the k-grid (applied with the $2\times2\times2$ supercell) all fold-down on the Gamma point. For computational tractability we slightly lowered the plane-wave cut-off from 350 to 300~eV and retrained the $2\times2\times2$ cell with the same cut-off for a fair comparison. The smaller plane-wave basis results in higher Pulay stress, which slightly affects the volume, but does not qualitatively change the crystal structure\cite{Lahnsteiner:prm18}.

Figure~\ref{fig:sizedep} shows $g_{\rm inorganic}(r)$ and $g_{\rm C,N}(r)$ for the standard supercell and the eight times larger one. The $g_{\rm inorganic}(r)$ of the two systems are almost identical. The main deviations in $g_{\rm C,N}(r)$ are the result of a different temperature at which the system switches from the Ort to Tet phase. This is to be expected for simulations with finite system size. The MD trajectory is chaotic and the transition does not occur at the same temperature even when the initial conditions are the same. This very good agreement indicates that the applied $2\times2\times2$ supercell is large enough to capture the crystal symmetry. This is in agreement with a fully \textit{ab-initio} MD study of the system size dependence of MAPbI$_3$ going up to the $6\times6\times6$ supercell\cite{Lahnsteiner:prb16}. 

%-------------------------------------------------------------------------
\begin{figure}[!t]
    \begin{center}
    \includegraphics[width=\columnwidth ,clip=true]{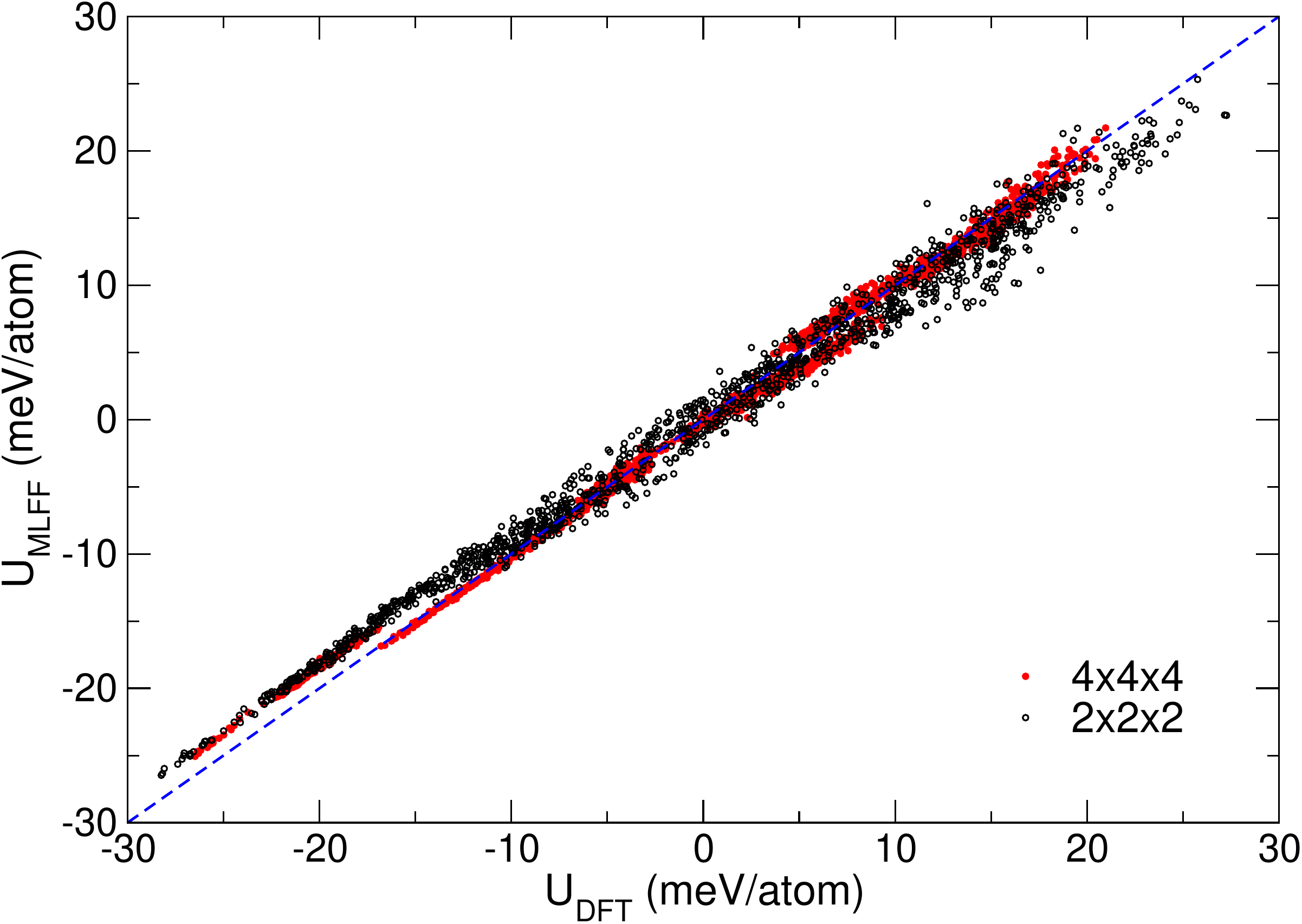}
    \end{center}
   \caption{DFT internal energies versus predicted MLFF energies for the structures in the MAPbBr$_3$ training-set obtained during on-the-fly heating MD in a $2\times2\times2$ and $4\times4\times4$ supercell.}
\label{fig:rms}
\end{figure}
%-------------------------------------------------------------------------

The accuracy of the MLFF model over the entire collected dataset of structures, which includes three different perovskite phases is very high. The DFT reference energy ($\rm U_{DFT}$) and predicted MLFF energy ($\rm U_{MLFF}$) for the test systems are plotted in Figure~\ref{fig:rms}, and show a clear linear relation over a large energy range. Note that both point clouds nicely overlap, whereby the larger variations are, as would be expected, observed for the smaller supercell. The overall root-mean-square (rms) error on the energy is only 1.7 and 0.88 meV/atom for the MLFF trained on the $2\times2\times2$ and $4\times4\times4$ supercells, respectively. This is small and of the same order of magnitude of state-of-the-art ML potentials (kernel-regression\cite{Bartok:prx18}, neural networks\cite{Behler:cmrev21}, etc.). Furthermore, for the two system sizes the errors in the force are 0.081 and 0.077 eV/\AA{} and in the stress 1.1 and 0.46 kB. These error estimates are typical for all MLFFs presented in this work. For instance, the rms errors in the energy of the six MLFFs ($X$=I,Br,Cl/sA or sB) are all in the 1.2-1.8 meV/atom range.

%-------------------------------------------------------------------------
\begin{figure}[!b]
    \begin{center}
    \includegraphics[width=\columnwidth ,clip=true]{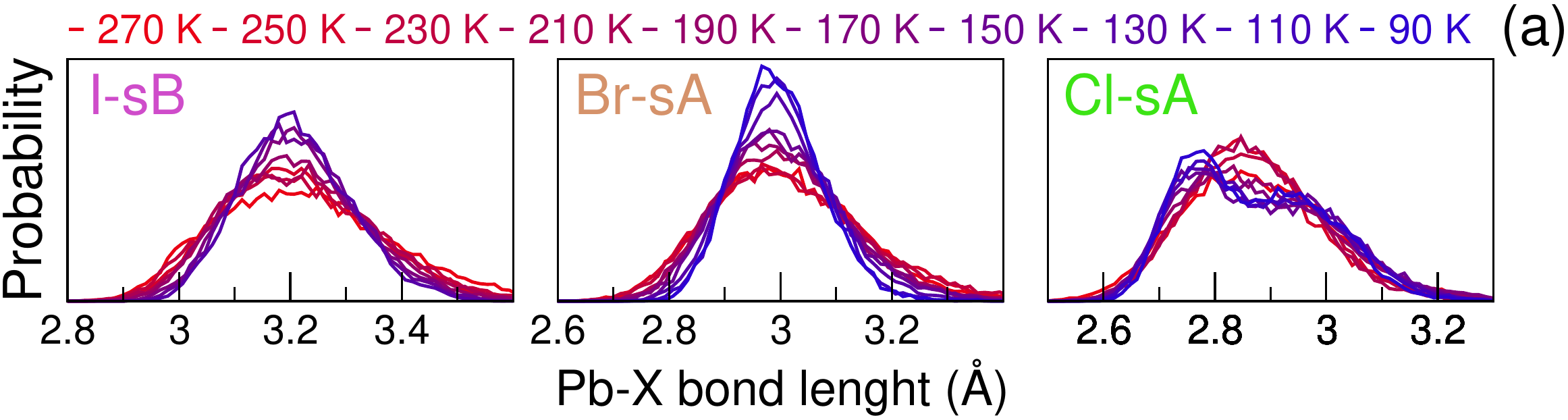}
    \includegraphics[width=\columnwidth ,clip=true]{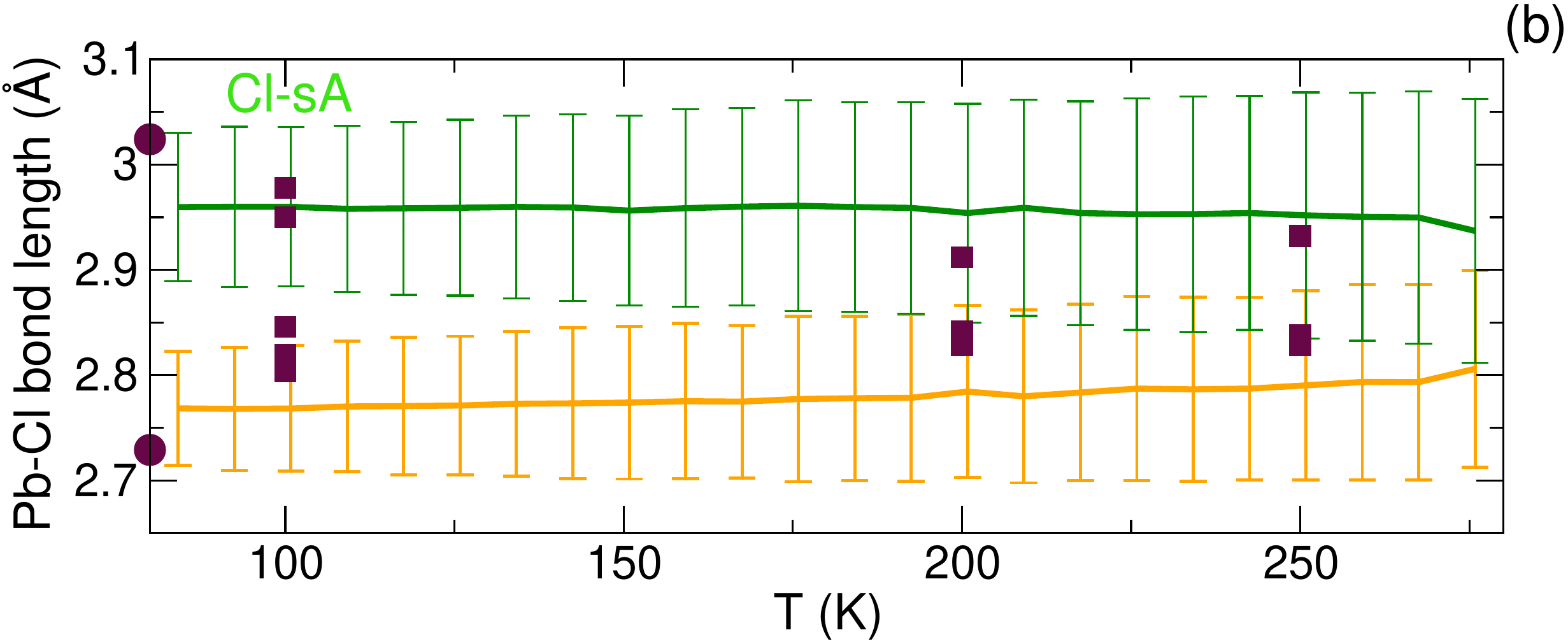}
    \end{center}
   \caption{(a) Distributions of Pb-$X$ bond lengths as function of temperature. (b) Mean values ($\mu$) and the standard deviations ($\pm\sigma$ errorbars) as function of temperature for the Pb-Cl bond lengths. The experimental values from Refs.~\cite{Chi:jssc05} and \cite{Bernasconi:jpcc18} are shown by the circles and squares, respectively.}
\label{fig:5}
\end{figure}
%-------------------------------------------------------------------------

\subsection{Octahedra distortions: dynamic or permanent?}
 The central question remains, why is the sA configuration more stable in MAPbCl$_3$ and to a less extent in MAPbBr$_3$ as compared to sB? Chi.\textit{et.al.} have already shown that the PbCl$_6$ octahedra are distorted\cite{Chi:jssc05}. This \textit{polar} distortion is highlighted in Fig.~\ref{fig:1} by the green and orange lines indicating the difference between two Pb-Cl bond lengths (3.02 and 2.73~\AA) in the crystallographic $a$-direction, and is in good agreement with our simulations. However, we find no noticeable distortion of the PbI$_6$ and PbBr$_6$ octahedra above 80~K, as shown in Figure~\ref{fig:5}. Again, in agreement with experiments of Refs.~\cite{Govinda:jpcl16,Swainson:jssc03}, however, opposite to the findings of Refs.~\cite{Page:acie16,Bernasconi:acsenl17} where a permanent octahedra distortion is found in MAPbBr$_3$ around room temperature. In Fig.~\ref{fig:5}(a) the distribution of the Pb-$X$ bond lengths as function of temperature are shown. The heating trajectories were cut in parts of equal length and all bond lengths in the $a$-direction were added to the distribution. The low-temperature distribution for Pb-Cl has two peaks. This distortion is not observed when starting from the Cl-sB structure, nor when starting from the Br-sB or I-sA structures. The distribution is well described by a combination of two Gaussian distribution functions $\mathcal{N}(\mu_1,\sigma_1)+\mathcal{N}(\mu_2,\sigma_2)$. For Cl-sA, the mean values ($\mu$) and the standard deviations ($\sigma$) as function of temperature are shown in Fig.~\ref{fig:5}(b). These values have been obtained from a separate heating trajectory with the finished MLFF on a $4\times4\times4$ MAPbCl$_3$ supercell. The so obtained distributions agree with Fig.~\ref{fig:5}(a) and improve statistical accuracy. Experimentally determined bond lengths from Refs.\cite{Chi:jssc05} and \cite{Bernasconi:jpcc18} have been added to Fig.~\ref{fig:5}(b) and agree within the standard deviation. In the simulations a single Gaussian suffices above $\sim$175~K, ie. $|\mu_1-\mu_2|< (\sigma_1+\sigma_2)$. At this temperature, the octahedral polar distortion is no longer observed on \textit{time average} and the crystal is in the cubic phase. As for I and Br, \textit{instantaneous} distortions of the octahedra do occur at these elevated temperatures.

The following scenario now becomes plausible, the sA ordered molecules are stabilized in the MAPbCl$_3$ orthorhombic phase by an anti-ferroelectric stripes pattern of dipolar octahedra in the $a$-direction. As argued in Refs.~\cite{Chi:jssc05} the volume of the perovskite has to be sufficiently small to induce these distortions, whereby the 'hard' MA deforms the 'soft' octahedra. However, the stripes pattern cannot be the only stabilization mechanism, because no distortion is observed in Br-sA.\newline{}

%-------------------------------------------------------------------------
\begin{figure}[!t]
    \begin{center}
    \includegraphics[width=\columnwidth ,clip=true]{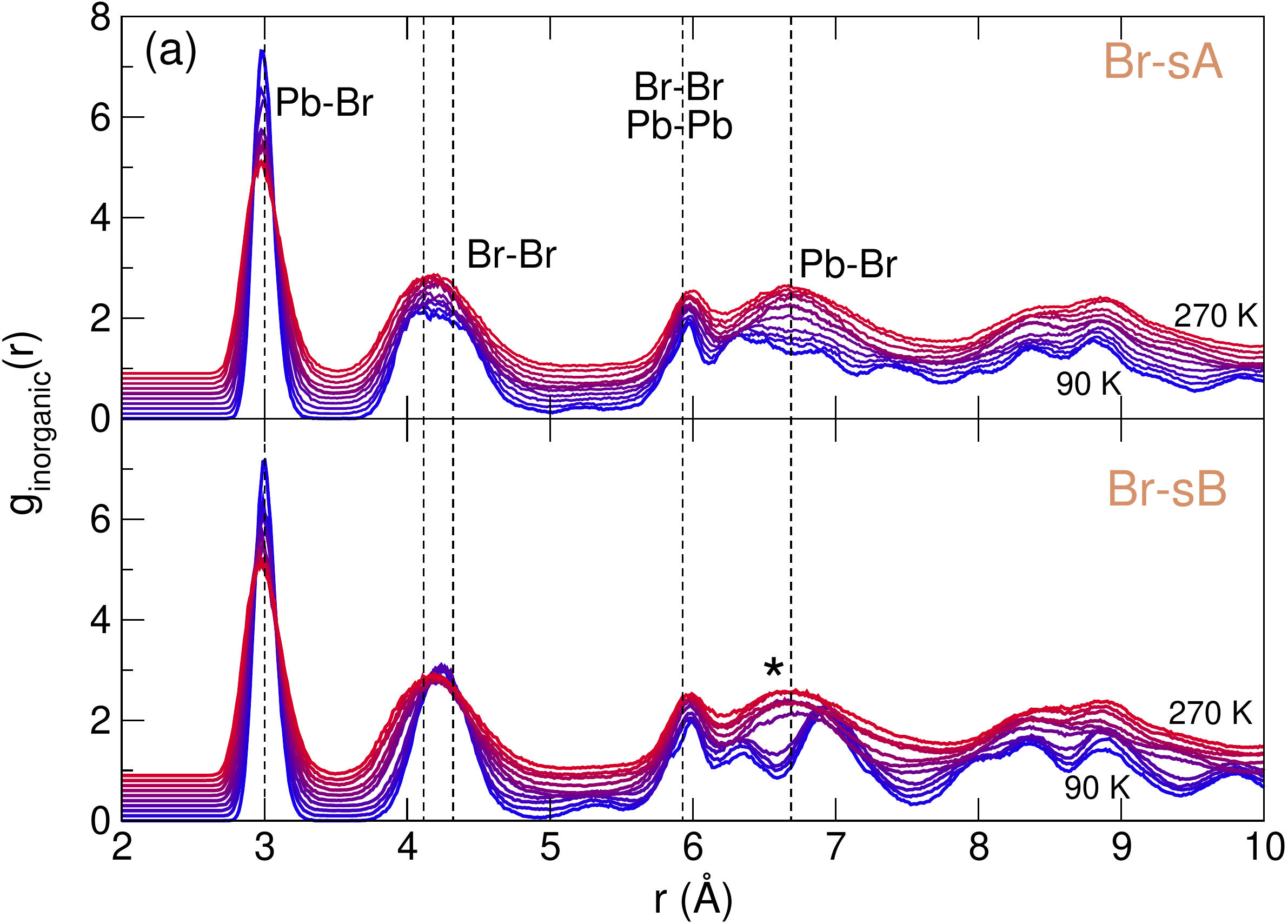}
    \includegraphics[width=\columnwidth ,clip=true]{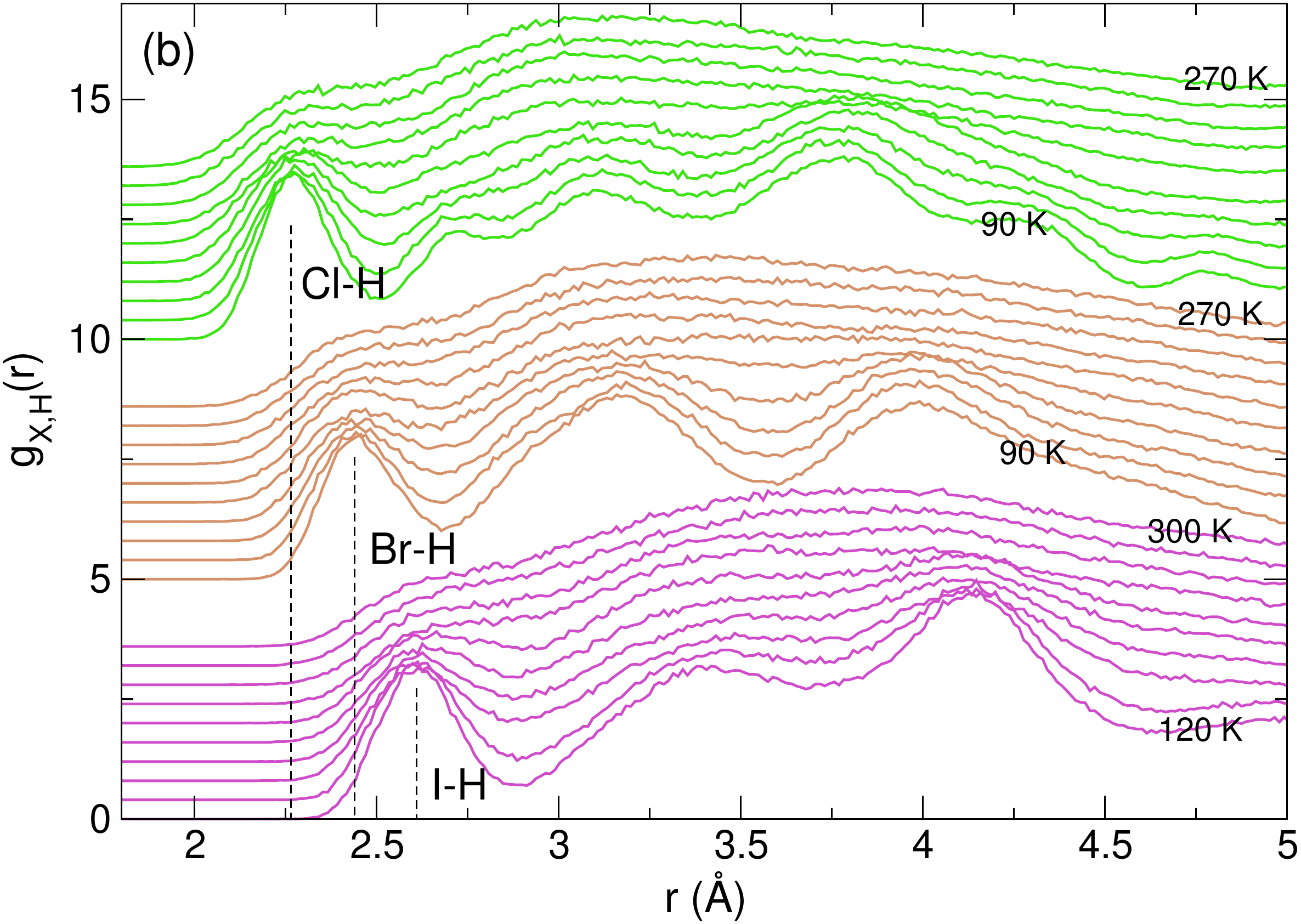}
    \end{center}
   \caption{Pair distribution functions of the (a) inorganic components in MAPbBr$_3$ and (b) $g_{X{\rm ,H}}(r)$ for increasing temperatures with steps of 20~K. Vertical dashed lines in (a) show the peak positions in the X-ray PDF of Ref.~\cite{Bernasconi:acsenl17} and (b) the first peaks in the calculated $g_{X{\rm ,H}}(r)$.}
\label{fig:rdf}
\end{figure}
%-------------------------------------------------------------------------

We would like to note that our 'ensemble average' view on the structural model results in PDFs for MAPbBr$_3$ that qualitatively agree with those obtained from X-ray diffraction experiments of Ref.~\cite{Bernasconi:acsenl17}. In Figure~\ref{fig:rdf}(a) $g_{\rm inorganic}(r)$ has been plotted on the same length scale as Fig.~2 in Ref.~\cite{Bernasconi:acsenl17}, whereby the vertical dashed lines indicate the experimental peak positions. Starting from sA, we also do not observe any relevant structural change in the 150-280~K temperature range apart from thermal broadening of the peaks. However, our approach classifies tetragonal and cubic structures within this range, and does not indicate that an orthorhombic structure would be a better fit throughout this temperature range. Starting from sB results in structural changes (indicated by *) in disagreement with the PCF of Ref.~\cite{Bernasconi:acsenl17}. This is another indication that the sB pattern is not the stable low temperature structure. The different structural interpretation of the crystal at elevated temperatures becomes apparent in the H-$X$ bonding as shown by $g_{{\rm H},X}(r)$ in Fig.~\ref{fig:rdf}(b). We find a halogen dependent first peak position. This finding is different from the 2.5~\AA{} peak observed in the neutron PDFs of both MAPbBr$_3$ and MAPbCl$_3$\cite{Bernasconi:jpcc18}.
\newline{}

\subsection{MA $C_3$ dynamics}
%-------------------------------------------------------------------------
\begin{figure}[!b]
    \begin{center}

    \includegraphics[width=0.75\columnwidth ,clip=true]{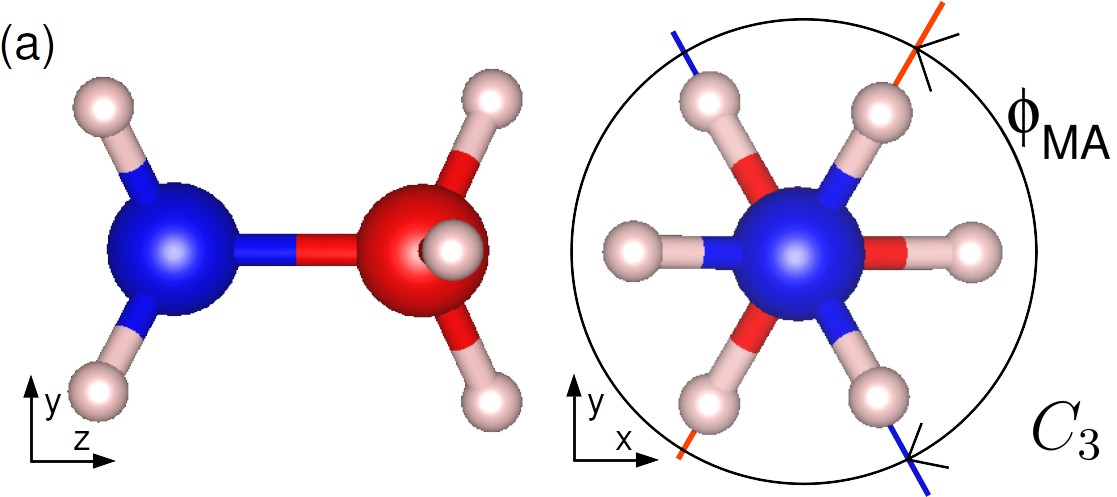}
   \includegraphics[width=\columnwidth ,clip=true]{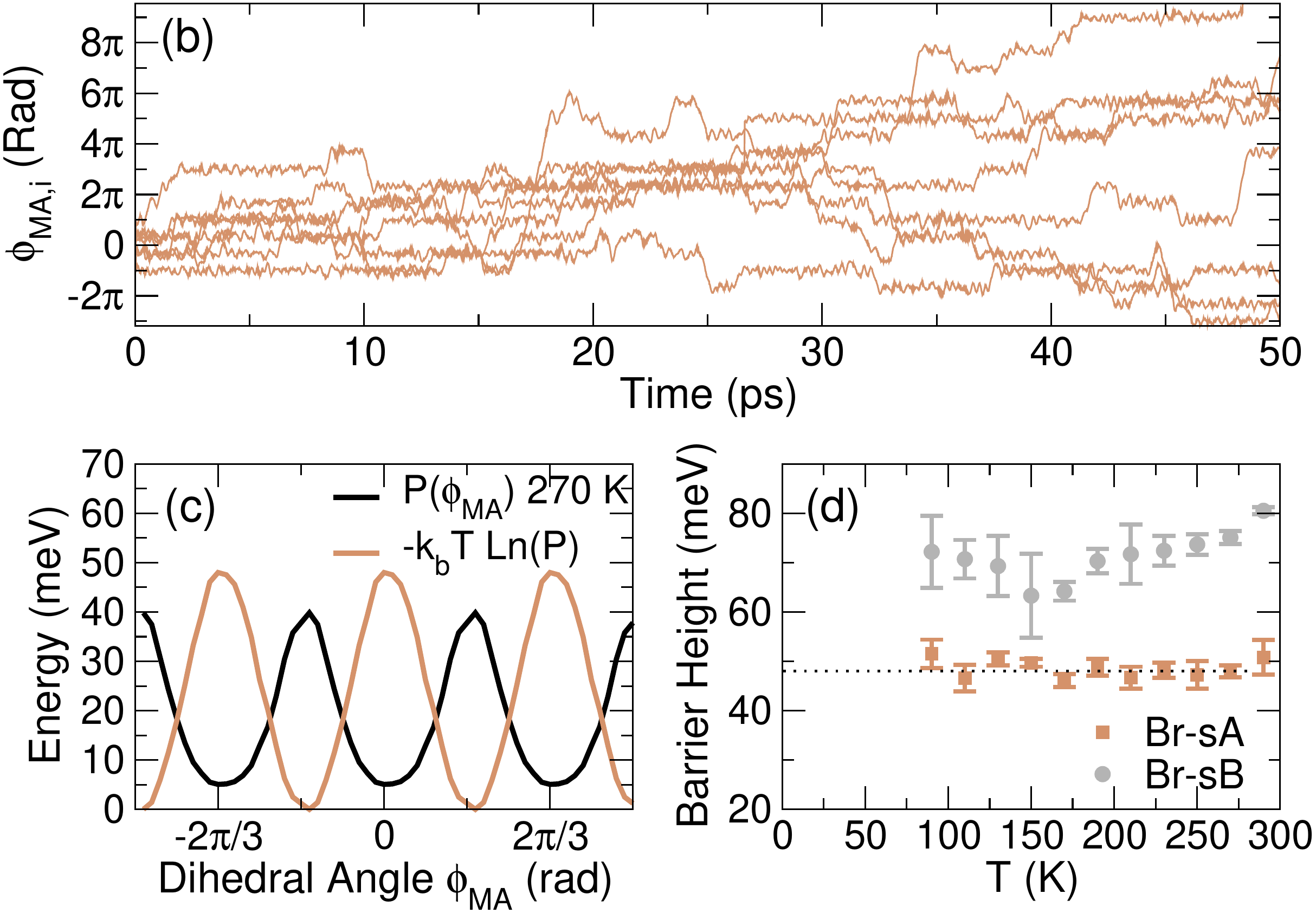}
    
    \end{center}
   \caption{3-fold dynamics of the NH$_3$ versus the CH$_3$ group. (a) Sketch of the angle $\phi_{\rm MA}$ in the MA molecule. (b) $\phi_{{\rm MA},i}$ as function of time for each of the eight molecules at 270~K. (c) Dihedral potential obtained by a Boltzmann inversion of the distribution (P) at 270~K. (d) Barrier heights as function of temperature. }
\label{fig:6}
\end{figure}
%-------------------------------------------------------------------------

We would like to note that training of a very accurate MLFF for the hyrbid perovskites is not fully completed by the here performed single heating run. Precise values for phase transition temperatures, $c/a$ ratios, etc., were not the aim of this work, but can be obtained with an accurate MLFF which enables long MD trajectories on large supercells\cite{Jinnouchi:prl19}. Limits to the accuracy can be seen in the $C_3$ \textit{torsion/rotation} degree of freedom of the molecule, for example. Figure~\ref{fig:6} shows the NH$_3$ versus the CH$_3$ group dihedral angle ($\phi_{\rm MA}$) of all MAs in MAPbBr$_3$ in $NPT$ ensembles with the finished MLFF. \textit{Torsion} ($\phi_{\rm MA}< 60^{\rm o}$) unfreezes at $\sim25$~K\cite{Sharma:jpcl20} and also \textit{rotations} ($\phi_{\rm MA}> 60^{\rm o}$) occur around our starting temperature (80~K) of the on-the-fly training.

Fig.~\ref{fig:6}~(b) shows that $\phi_{{\rm MA},i}(t)$ of a single molecule shows step-like behavior, occasionally jumping between planes separated by 120$^{\rm o}$, and is superimposed by a fast oscillation. For each of the eight molecules in the supercell a probability distribution of $\phi_{\rm MA}$ as in Fig.~\ref{fig:6}(c) was made. We then calculate the $C_3$ rotational energy barrier in the two MLFFs of MAPbBr$_3$ (sA and sB) by a Boltzmann inversion of the distribution. A barrier was only assigned at a temperature $T$ when the number of 120$^{\rm o}$ rotations in the $\sim100$~ps long MD trajectories exceeded the number of molecules in the supercell. The error bars in Fig.~\ref{fig:6}(d) correspond to $\pm\sigma$, the standard deviation of the eight obtained barriers. The barrier is, within our statistical accuracy, temperature independent and, surprisingly, different between sA and sB. It is tempting to conclude that the sA structure for Br affords a more facile $C_3$ rotational degree of freedom compared to the sB structure within the orthorhombic phase. However, the difference should not persist in the high temperature cubic phase, in which the MAs are orientationally disordered. This should be a warning that training is not yet completed and the MLFF still shows a bias depending on the initial conditions.

We are able to measure the barriers of a single molecule in vacuum in the same manner. The absence a surrounding Pb-Br framework does not destabilize the molecule. Barriers obtained in this way are slightly lower than the DFT value of 105~meV for MA in vacuum. This value was calculated as the difference in internal energies of the optimized $\phi_{\rm MA}=0^{\rm o}$ and $60^{\rm o}$ molecule, for which all internal degrees of freedom were relaxed under the constraint of $\phi_{\rm MA}$. Using these two structures the barriers for the Br-sA and Br-sB MLFFs are 78 and 89~meV, respectively.

Since both Br-sA and Br-sB show $C_3$ rotations at temperatures below the Ort-Tet phase transition temperature, we can conclude that the unfreezing of this motion does not drive it. This is in agreement with the findings of the computational study of Kieslich~\textit{et.al.}\cite{Kieslich:com18}. This phase transition is driven by an increase in $C_4$ configurational entropy of the molecules\cite{Onoda-Yamamuro:jpcs90}. This, however, does not preclude the possibility that entropy related to $C_3$ dynamics is involved in this phase transition. Based on the observed initial condition dependence (sA or sB) of the barrier, we \textit{speculate} that Br-sA is entropically stabilized by the $C_3$ torsion/rotation degree of freedom of the molecule, and thereby determines the orthorhombic structure below the phase transition temperature.

\section{Conclusions}

In conclusion, we have shown that \textit{on-the-fly} machine-learning force fields are a very powerfull tool in determining atomic structure in dynamic, entropically stabilized solids. Already during the training-by-heating MD important structural characteristics are qualitatively correct and even quantitatively useful. As a prime example, the low-temperature ordering pattern of MA molecules in MAPb$X_3$ perovskites, which is not uniquely resolved by diffraction experiments, was studied. We determined the most likely structure by slow-heating DFT-based molecular dynamics and analyzing the \textit{librational pathways}. By comparing this analysis with reported temperature dependent lattice parameters and refined structures, we show that the ordering of the molecules (sA or sB) in orthorhombic phases of MAPbBr$_3$ and MAPbCl$_3$ is similar (sA), while in MAPbI$_3$ they are differently ordered (sB). This is unexpected, since the sA pattern is energetically unfavorable when considering solely the intrinsic dipole moment of the MA molecules. The sA order induces a permanent structural distortion of the PbCl$_6$ octahedra at low temperature, resulting in an anti-ferroelectric stripes pattern in the crystallographic $a$-direction. In the higher temperature cubic phase this distortion is no longer observed in the ensemble average, instead instantaneous dynamic distortions appear. No permanent distortion is observed in the PbBr, nor PbI, octahedra even down to the lowest simulated temperature of 80~K. We have presented indications that the sA order in low temperature, orthorhombic MAPbBr$_3$ is stabilized by an entropic contribution to the free energy related to the $C_3$ dynamics of the MA molecules. We hope that this paper will stimulate combined experimental and MLFF studies of the structure of many other complex Dynamic Solids.

\begin{acknowledgement}
M.B. and J.L. gratefully acknowledge funding by the Austrian Science Fund (FWF): P 30316-N27. Computations were partly performed on the Vienna Scientific Cluster VSC3 and on the Dutch national e-infrastructure with the support of SURF Cooperative. 
\end{acknowledgement}

\begin{suppinfo}
Supporting information can be downloaded from the website of the Journal of Physical Chemistry C (2021): \href{https://doi.org/10.1021/acs.jpcc.1c06835}{10.1021/acs.jpcc.1c06835}
\begin{itemize}

\item For each of the six on-the-fly heating trajectories a movie was generated. A running average (window size 25~K) over each atomic coordinate is computed to smoothen out high frequency movements. At high temperature, where the molecules rotate fast, this contracts the atoms in the molecule to a point. The video shows the $2\times2\times2$ supercell under periodic boundary conditions and from two perspectives (from the top and side).

\item The DFT-SCAN optimized (0~K) orthorhombic sA and sB structures for MAPbI$_3$, MAPbBr$_3$ and MAPbCl$_3$ in .cif file format.
  
\end{itemize}
\end{suppinfo}

\providecommand{\latin}[1]{#1}
\makeatletter
\providecommand{\doi}
  {\begingroup\let\do\@makeother\dospecials
  \catcode`\{=1 \catcode`\}=2 \doi@aux}
\providecommand{\doi@aux}[1]{\endgroup\texttt{#1}}
\makeatother
\providecommand*\mcitethebibliography{\thebibliography}
\csname @ifundefined\endcsname{endmcitethebibliography}
  {\let\endmcitethebibliography\endthebibliography}{}

\end{document}